\documentclass[twocolumn,showpacs,preprintnumbers,amsmath,amssymb]{revtex4}

\usepackage{graphicx}
\usepackage{dcolumn}
\usepackage{bm}
\usepackage{epsfig}

\begin{document}

\preprint{APS/123-QED}

\title{$^{70}$Ge(p,$\gamma$)$^{71}$As and $^{76}$Ge(p,n)$^{76}$As cross sections for the astrophysical $p$ process: sensitivity of the optical proton potential at low energies}

\author{G. G.\,Kiss}%
 \email{ggkiss@atomki.hu}
\author{Gy.\,Gy\"urky}%
\author{Z.\,Elekes}%
\author{Zs.\,F\"ul\"op}%
\author{E.\,Somorjai}%
\affiliation{%
Institute of Nuclear Research (ATOMKI), H-4001 Debrecen, POB.51., Hungary}%
\author{T.\,Rauscher}%
\affiliation{%
Universit\"at Basel, CH-4056 Basel, Switzerland}%

\author{M.\,Wiescher}%
\affiliation{%
University of Notre Dame, Notre Dame, Indiana 46556, USA}%

\date{\today}

\begin{abstract}
The cross sections of the
$^{70}$Ge(p,$\gamma$)$^{71}$As and $^{76}$Ge(p,n)$^{76}$As
reactions have been measured with the activation method in the Gamow window for the astrophysical $p$ process. The experiments were carried out at the Van de Graaff and
cyclotron accelerators of ATOMKI. The cross sections have been
derived by measuring the decay $\gamma$-radiation of the reaction
products. The results are compared to the predictions of
Hauser-Feshbach statistical model calculations using the code
NON-SMOKER. Good agreement between theoretical and experimental
$S$ factors is found. Based on the new data, modifications of the optical potential
used for low-energy protons are discussed.

\end{abstract}

\pacs{25.40.Lw, 26.30.+k, 26.50.+x, 27.50.+e}%

\maketitle

\section{Introduction}

The stable proton-rich nuclei with charge number Z$\geq$34 are
called $p$ nuclei \cite{woosley}. The natural isotopic abundance of
these nuclei is $10-100$ times less than the more neutron-rich
isotopes. Most $p$ nuclei cannot be produced by neutron capture
because they are separated by unstable short-lived nuclei from the
$s$ or $r$ process path. It is generally accepted that the main
stellar mechanism synthesizing these nuclei -- the so-called
$p$ process -- is initiated by ($\gamma$,n) photodisintegration
reactions on preexisting more neutron-rich seed nuclei. As the
neutron separation energy increases along this path towards more
neutron deficient isotopes, ($\gamma$,p) and ($\gamma$,$\alpha$)
reactions become stronger and process the material towards lower
masses \cite{arnould,rauscher06,rapppn}. Despite considerable
experimental and theoretical efforts in recent years, there are
still open questions about the nature of the $p$ process and the
synthesis of the $p$ isotopes. 

The high intensity energetic photons
necessary for these $\gamma$-induced reactions are available only
in scenarios with temperatures around 2-3 GK.
One possible site for such a scenario is the O-Ne layer of a
massive star in hydrostatic pre-supernova burning or explosive burning
due to the type II supernova shockwave. Other
potential sites providing the necessary conditions for $p$ process
nucleosynthesis have been summarized recently \cite{arnould}. One
of the main uncertainties in $p$ process nucleosynthesis is
associated with the origin of the light Mo, Ru, In, and Sn
p-nuclei with a fairly large abundance which cannot be explained
in the framework of standard $p$ process nucleosynthesis models. It
has been argued that the $p$ process might be complemented by other
nucleosynthesis processes such as the $rp$ process for the light $p$ nuclei
(A$\leq$100) \cite{schatz} or the neutrino induced $\nu p$ process
in type II supernovae \cite{frohlich}. Additional
possibilities are discussed in \cite{arnould}.

Modeling the synthesis of the $p$ nuclei and calculating their
abundances require an extended reaction network calculation
involving more than 10000 reactions on 2000 stable and unstable
nuclei. Most of the reaction rates are calculated by using the
Hauser-Feshbach statistical model. The rates of the
$\gamma$-induced reactions can be determined experimentally by
measuring the inverse reaction cross section and converting the
results by using the detailed balance theorem. In contrast to
neutron-capture reactions which are comparatively well studied
over the relevant mass region of the stable isotopes, charged
particle induced reactions at energies below the Coulomb barrier
are only scarcely studied for the mass region above iron. Previous
\cite{woosley} and
recent \cite{rauscher06,rapppn} investigations agree on the fact
that ($\gamma$,$\alpha$)
reactions are mainly important at higher masses while ($\gamma$,p)
reactions are more important for the lighter $p$ nuclei. In recent
years a range of ($\alpha$,$\gamma$) reaction cross sections on
$^{70}$Ge, $^{96}$Ru, $^{106}$Cd, $^{112}$Sn, and $^{144}$Sm have
been measured, and the results have been compared with model
predictions \cite{fuge, rappru, gyurkyag, ozag, som}. In general
the models were able to reproduce the experimental data within a
factor of two; in some cases, however, larger deviations have been
observed.

\begin{figure*}
\resizebox{2.0\columnwidth}{!}{\rotatebox{270}{\includegraphics[clip=]{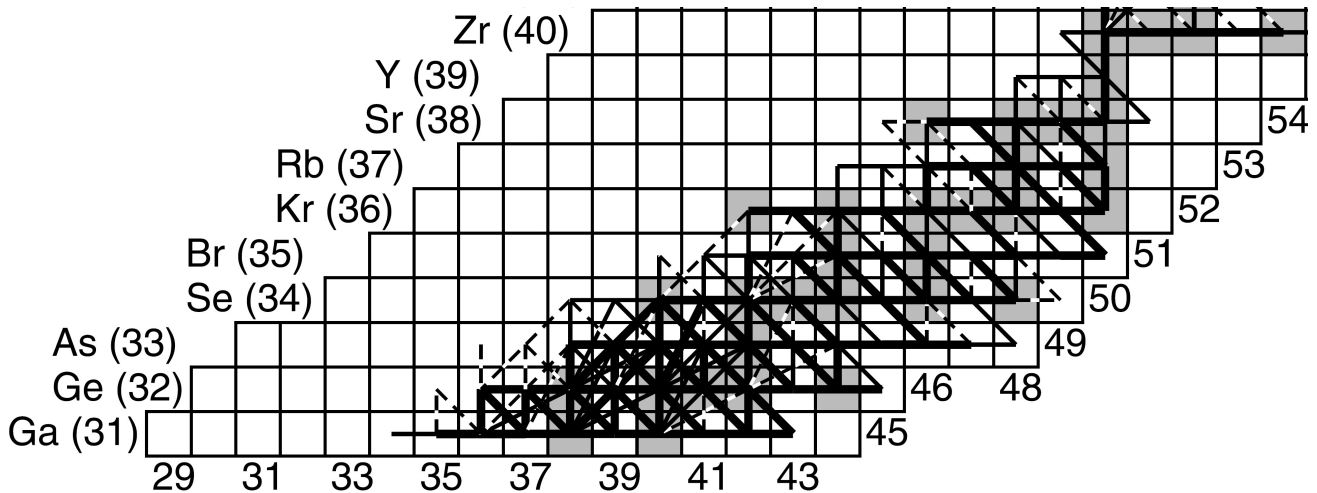}}}
\caption{\label{fig:reacpattern}Integrated reaction flux of the $p$ process in the Ga to Zr
mass range during the first second of a type II SN explosion
triggering a shock front in the Ne/O layer with a maximum
temperature of T\,=\,3\,GK. The strength of reaction flux is
indicated by line thickness. Figure is taken from \cite{rapppn}.}
\end{figure*}

Beside these studies, numerous (p,$\gamma$) experiments
have been performed on stable $p$ nuclei to determine the reaction
rates of the corresponding ($\gamma$,p) reactions \cite{laird,
sauter, bork, chloupek, gyusr, hariss, ozkan, galanopoulos,
gyuse, tsagari, gyucd}. These proton capture studies have shown
better agreement between experimental results and theoretical
predictions than in the case of alpha capture reactions discussed
above. However, the existing experimental database is not
sufficient for a global check of the reliability of model
calculations; further experimental data are clearly needed
\cite{rapppn, rauscher06}.

As will be further elaborated in the discussion section of this
paper, (p,n) reactions generally are better suited to study the impact
of the optical potential used for protons. While uncertainties in
both the proton and $\gamma$ widths impact the cross section prediction
in the experimentally accessible energy range of capture reactions,
the theoretical uncertainties in (p,n) reactions are dominated by the
proton width alone. Thus, (p,n) studies allow global testing of the proton
potential whereas (p,$\gamma$) ones directly study astrophysically
important reactions.

Moreover, it was demonstrated recently that (p,n) reactions on $s$ or
$r$ process seed nuclei do affect the abundances of the light
$p$ nuclei \cite{rapppn}. There is only limited experimental
information available about the low energy cross sections of
critical (p,n) reactions in this mass range. This is demonstrated
in the case of $^{75}$As(p,n)$^{75}$Se, where the existing data
\cite{Kail79,Mus88} show considerable discrepancies which makes a
direct comparison with Hauser Feshbach predictions difficult
\cite{rapppn}. A reliable simulation of the rather complex
$p$ process nucleosynthesis pattern in this mass range is of great
relevance in particular to differentiate between the contributions
from $p$ process, $s$ process, and $r$ process sites. This also requires
testing the applicability and reliability of the global Hauser
Feshbach predictions in this mass range.

In this paper we want to pursue our studies investigating the cross
sections of $^{70}$Ge(p,$\gamma$)$^{71}$As and $^{76}$Ge(p,n)$^{76}$As
which are both associated with the reaction flow pattern in this
mass range as demonstrated by \cite{rapppn}. Figure \ref{fig:reacpattern}
shows the complex reaction pattern during the first second of the
shock front induced $p$ process. The reaction
$^{70}$Ge(p,$\gamma$)$^{71}$As corresponds to a branch
associated with the final abundance of the
$p$ nucleus $^{70}$Ge in the $p$ process. The proton capture direction gives rise
to depletion of this nucleus at low $p$ process temperatures whereas the
reverse rate can produce it through a ($\gamma$,n)-($\gamma$,p) branching at higher
temperature. The reaction
$^{76}$Ge(p,n)$^{76}$As is directly associated with the
transformation of the $r$ process seed nucleus $^{76}$Ge to the
$p$ nucleus $^{74}$Se in the first moments of $p$ process
nucleosynthesis; for example through the
$^{76}$Ge(p,n)$^{76}$As($\gamma$,n)$^{75}$As(p,n)$^{75}$Se($\gamma$,n)$^{74}$Se
reaction chain.  Fig.\ \ref{fig:reacpattern} suggests that many alternative
reaction sequences exist, but it also underlines that
during the $p$ process $^{76}$Ge is not only depleted but also
produced by an initial
$^{74}$Ge(n,$\gamma$)$^{75}$Ge(n,$\gamma$)$^{76}$Ge reaction flux.
This modifies the initial $^{76}$Ge abundance. A detailed
knowledge of the reaction rate of $^{76}$Ge(p,n)$^{76}$As as the
sole depletion process is important to investigate the overall
nucleosynthesis pattern and history of $^{76}$Ge in a type II
supernova shock front environment.

\begin{table}
\caption{\label{tab:decay}Decay parameters of $^{70}$Ge(p,$\gamma$)$^{71}$As and $^{76}$Ge(p,n)$^{76}$As reaction
products taken from the literature.}
\setlength{\extrarowheight}{0.1cm}
\begin{ruledtabular}
\begin{tabular}{ccccc}
\parbox[t]{1cm}{\centering{Product \\ nucleus}} &
\parbox[t]{1.2cm}{\centering{Half-life [hour]}} &
\parbox[t]{1.8cm}{\centering{Gamma \\energy [keV]}} &
\parbox[t]{1.6cm}{\centering{Relative \\ $\gamma$-intensity \\ per decay [\%]}} &
\parbox[t]{1.cm}{\centering{Ref.}} \\
\hline
$^{71}$As &  65.28 $\pm$ 0.15 & 174.95 $\pm$ 0.04 & 82.00 $\pm$ 0.25 & \cite{NNDC1} \\
& & 326.79 $\pm$ 0.02 & 3.03 $\pm$ 0.03  \\
& & 499.91 $\pm$ 0.01 &  3.62 $\pm$ 0.02\\
& & 1095.51 $\pm$ 0.01 &  4.08 $\pm$ 0.06\\ \\
$^{76}$As & 25.87 $\pm$ 0.05 & 559.10 $\pm$ 0.01 & 44 $\pm$ 1  & \cite{NNDC2} \\
& & 657.04 $\pm$ 0.01 & 6.2 $\pm$ 0.3 &  \\
\end{tabular} \label{decaypar}
\end{ruledtabular}
\end{table}

In the following we outline the experimental approach for
measuring $^{70}$Ge(p,$\gamma$)$^{71}$As and $^{76}$Ge(p,n)$^{76}$As
using the activation method. We describe the experimental
set-up and procedure and finally present the
experimental results. A concluding discussion section contains
a detailed theoretical analysis of the results and their importance for
the prediction of low-energy optical potentials.

\section{Investigated reactions}

To determine the cross sections we used the activation method. This method allows to measure several cross sections simultaneously using natural targets. The element Ge has five stable isotopes with mass numbers A=70,
72, 73, 74, and 76, having isotopic abundances of 20.37\%, 27.31\%,
7.76\%, 36.73\%, and 7.83\%, respectively \cite{NNDC3}. The number of reaction channels measurable with the activation method is limited. Specifically, because of the low (p,n) thresholds the $^{73,74}$Ge(p,n) reactions cannot be separated from the $^{72,73}$Ge(p,$\gamma$) ones.
It is also impossible to measure the cross sections of the $^{74}$Ge(p,$\gamma$)$^{75}$As
and $^{76}$Ge(p,$\gamma$)$^{77}$As reactions because $^{75}$As is stable
and in the case of the second reaction the $\gamma$-intensity from the decay of the
final nucleus is very weak. The $^{70,72}$Ge(p,n) reaction channels are not open in the investigated energy region.

In summary, it proved feasible to measure the proton induced cross sections of the
reactions $^{70}$Ge(p,$\gamma$)$^{71}$As and $^{76}$Ge(p,n)$^{76}$As
in the energy range $E_{c.m.}=1.6-4.3$ MeV. This energy range covers the Gamow window for typical $p$ process temperatures. The decay parameters used for the analysis are summarized
in Table \ref{tab:decay}.

\section{Experimental procedure}

\subsection{Target properties}

\begin{figure}
\resizebox{1.0\columnwidth}{!}{\rotatebox{270}{\includegraphics[clip=]{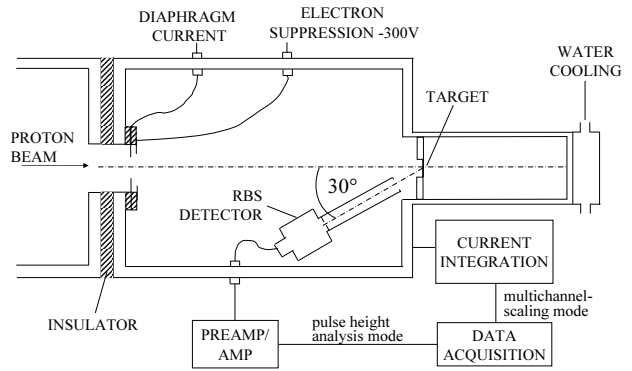}}}
\caption{\label{fig:chamber}Schematic view of the target chamber used for the irradiations}
\end{figure}

The targets were made by evaporating natural metallic Ge on thin,
high purity Al foils. Aluminum backings were used because Al has only one
stable isotope. The reaction product of
$^{27}$Al(p,$\gamma$)$^{28}$Si is stable and the (p,n) channel
opens only at $E_{c.m.}=5.59$ MeV which is above our investigated
energy range. Moreover, Al can be easily distinguished from Ge in the
Rutherford Backscattering (RBS) spectrum that was used to monitor the
target stability, see below. These reasons make Al ideal as
backing material in low-energy proton-induced reactions using the
activation technique. The thickness of the targets has been
derived by weighing. The weight of the Al backing was measured before and after the Ge evaporation. We repeated the mass measurement of the target also after the irradiation to prove that no changes in the number of the target atoms occurred during the irradiation.

The thickness of
the targets varied between 50 and 290 $\mu$g/cm$^2$,
corresponding to a proton energy loss of $\approx$ 2 keV ($E_\mathrm{p}=4.4$ MeV)
and $\approx$ 25 keV ($E_\mathrm{p}=1.6$ MeV), respectively. The proton
energy loss was calculated using the SRIM \cite{srim} code. Thicker targets
were used at low bombarding energy, where the cross section and
the corresponding $\gamma$-yield is small. Even in the case of the thickest
target at the lowest proton energy, the energy loss was $\approx$25keV
which is small compared to the 400 keV energy steps (see below).

\subsection{Irradiation and activity determination}

The irradiations were carried out at the Van de Graaff and
cyclotron accelerators of ATOMKI. The energy range
($E_\mathrm{p}=1.6$ to 4.4 MeV) was covered with 400 keV steps. A schematic
view of the target chamber is shown in Fig.\ \ref{fig:chamber}. Downstream the
diaphragm the whole chamber served as a Faraday cup. The collected
charge was measured with a current integrator. At the entrance of
the chamber another diaphragm was used with $-300$V secondary
electron suppression voltage.

\begin{figure}
\resizebox{1.0\columnwidth}{!}{\rotatebox{270}{\includegraphics[clip=]{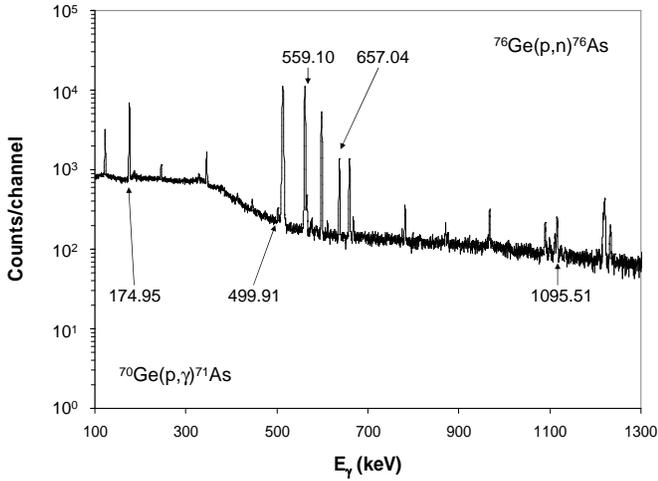}}}
\caption{\label{fig:spectrum}Off-line $\gamma$ spectrum taken after the irradiation
using 3.2 MeV proton beam. The marked $\gamma$ peaks were used for
the analysis. The other peaks correspond to
laboratory background or beam induced background on impurities in
the target or the backing.}
\end{figure}

Each irradiation lasted between 3 and 11 h. Before the experiment several beam tests were performed to verify target stability. These tests showed that there was no deterioration of the targets using a proton beam current less than 500 nA. A surface barrier detector was built into the chamber
at $\Theta$=150$^\circ$ relative to the beam direction. This
detector was used to monitor the target stability during the irradiation.
The RBS spectra were taken continuously and stored regularly during
the irradiation.

The collected charge varied between $\approx\,5-20\,\times\,10^{-3}$\,C in the case of each irradiation. The current integrator counts were recorded in multichannel scaling mode,
stepping the channel in every 10 s to take into account the
possible changes in the beam current.

Between the irradiation and the $\gamma$ counting, a waiting time of
0.5 h was inserted in order to decrease the yield of the disturbing
short-lived activities. The $\gamma$ radiation following the
$\beta$ decay of the produced As isotopes was measured with a
40\% relative efficiency HPGe detector. A 10 cm thick lead shield
was used to reduce the laboratory background.
The $\gamma$ spectra were taken for 10 h and stored regularly in
order to follow the decay of the different reaction products.

The absolute efficiency curve of the detector was measured using calibrated
$^{137}$Cs, $^{60}$Co, and $^{152}$Eu sources in the same geometry.
The measured points were fitted with a third-degree logarithmic
polynomial to determine the efficiency curve for the energy region
of interest. The efficiency of the detector was also studied with Monte Carlo
simulations using the GEANT code \cite{geant}. Good agreement was found between the efficiency curve calculated with GEANT and the polynomial fit.  As an example, Fig.\ \ref{fig:spectrum} shows a 
collected off-line $\gamma$ spectrum after irradiation using a 3.2 MeV proton beam.
\begin{figure}
\resizebox{0.9\columnwidth}{!}{\rotatebox{270}{\includegraphics[clip=true]{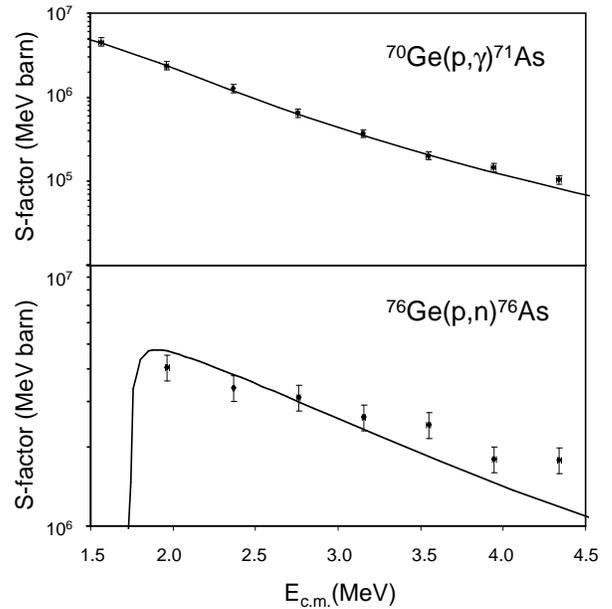}}}
\caption{\label{fig:expresults}Measured $S$ factor of the $^{70}$Ge(p,$\gamma$)$^{71}$As and $^{76}$Ge(p,n)$^{76}$As reactions compared to previous NON-SMOKER results \cite{adndt,adndtcs}.}
\end{figure}

\section{Experimental results}

Table \ref{tab:sfactors} summarizes the experimental cross sections and $S$ factors
for the two investigated reactions. Figure \ref{fig:expresults} shows
the results in comparison to the Hauser-Feshbach statistical
model prediction, using the NON-SMOKER code \cite{nonsmoker} with standard settings
as given in \cite{adndt,adndtcs}.
The uncertainties of the
center-of-mass energies given in the second column of Table \ref{tab:sfactors}
correspond to the energy loss in the target calculated with the SRIM
code and for the energy stability of the beam. The uncertainty in the cross section ($S$ factor) values is the
quadratic sum of the following partial uncertainties: efficiency of the
HPGe detector ($\approx$ 7\%), number of target atoms ($\approx$
6\%), current measurement ($\approx$ 3\%), uncertainties of the
level parameters found in literature \cite{NNDC1, NNDC2}  ($\leq$ 1\%), and counting
statistics (0.8 to 11\%).
\begin{table}
\caption{\label{tab:sfactors}Experimental cross sections and $S$ factors of the
$^{70}$Ge(p,$\gamma$)$^{71}$As and $^{76}$Ge(p,n)$^{76}$As reactions.}

\begin{ruledtabular}
\setlength{\extrarowheight}{0.1cm}
\begin{tabular}{lr@{\hspace{0.15cm}$\pm$\hspace{-0.25cm}}lr@{\hspace{0.15cm}$\pm$\hspace{-0.25cm}}
lr@{\hspace{0.15cm}$\pm$\hspace{-0.25cm}}l} $E_{\rm beam}$ &
\multicolumn{2}{c}{\hspace{-0.4cm}$E_{\rm c.m.}^{\rm eff}$} &
\multicolumn{2}{c}{\hspace{-0.4cm}Cross section} &
\multicolumn{2}{c}{\hspace{-0.5cm}$S$ factor} \\
{[keV]} & \multicolumn{2}{c}{\hspace{-0.2cm}[keV]} & \multicolumn{2}{c}{\hspace{-0.4cm}[mb]} &
\multicolumn{2}{c}{\hspace{-0.5cm}[10$^{6}$ MeV\,b]} \\
\hline
\\
\multicolumn{7}{c}{$^{70}$Ge(p,$\gamma$)$^{71}$As}\\ \\
1600 & 1565 & 13 & 0.035 & 0.004 & 4.5 & 0.5 \\
2000 & 1963 & 13 & 0.22 & 0.03 & 2.4 & 0.3 \\
2400 & 2363 & 4 & 0.71 & 0.08 & 1.3 & 0.1 \\
2800 & 2757 & 9 & 1.4 & 0.2 & 0.65 & 0.07 \\
3200 & 3152 & 10 & 2.4 & 0.3 & 0.37 & 0.04 \\
3600 & 3545 & 12 & 3.2 & 0.4 & 0.20 & 0.02 \\
4000 & 3942 & 12 & 4.9 & 0.6 & 0.15 & 0.02 \\
4400 & 4336 & 13 & 6.8 & 0.8 & 0.11 & 0.01 \\ \\
\multicolumn{7}{c}{$^{76}$Ge(p,n)$^{76}$As} \\ \\
2000 & 1965 & 13 & 0.37 & 0.04 & 4.0 & 0.4 \\
2400 & 2366 & 4 & 1.9 & 0.2 & 3.4 & 0.4 \\
2800 & 2760 & 9 & 6.8 & 0.8 & 3.1 & 0.3 \\
3200 & 3156 & 10 & 17 & 2 & 2.6 & 0.3 \\
3600 & 3549 & 12 & 40 & 4 & 2.4 & 0.3 \\
4000 & 3946 & 12 & 62 & 7 & 1.8 & 0.2 \\
4400 & 4341 & 13 & 115 & 13 & 1.8 & 0.2 \\
\end{tabular} \label{tab_ag}
\end{ruledtabular}

\end{table}

\section{Discussion}
\subsection{Comparison to theory}
The proton capture data is excellently predicted by the previously published
NON-SMOKER Hauser-Feshbach calculation \cite{adndt,adndtcs}. Also the (p,n) data is
reproduced reasonably well although the data seem to indicate a slightly different
slope of the $S$ factor as function of energy. Since the computation of the relevant
quantity for astrophysics, the reaction rate, involves an integration across the
energy range given by the Gamow peak (see, e.g., \cite{ili07}), the observed deviation is
averaged out and barely appears in the rates at $p$-process temperatures.
Nevertheless, it is worthwhile to study the origin of the different behavior of the
relation between data and predictions in the two cases of radiative capture and (p,n) reaction.

At first, it has to be realized that the theoretical (p,$\gamma$) and (p,n) results
exhibit different dependences on nuclear inputs. The averaged statistical model
cross sections $\left< \sigma \right>_\mathrm{HF}$ are derived from averaged
widths $<\Gamma>$, leading to an expression similar to that of resonant
reactions \cite{descrau}
\begin{equation}
\label{eq:hf}
\left< \sigma \right>_\mathrm{HF} \propto \frac{<\Gamma>_i<\Gamma>_o}{<\Gamma>_\mathrm{tot}} \quad ,
\end{equation}
with $i$, $o$ labeling the entrance and exit channel, respectively, and $<\Gamma>_\mathrm{tot}$ being the averaged
total width including all possible deexcitation channels (particle and
photon emission) of the compound nucleus. As is well-known in the resonant case,
the energy dependence is then determined by the smallest width in the numerator of
Eq.\ (\ref{eq:hf}). The situation is more involved if the widths are of similar
magnitude or if another channel is significantly contributing to the total width.

In order to study the sensitivities of the calculated $S$ factors, we systematically
and independently  varied the averaged widths of the neutron, proton, and $\gamma$
channels by factors of
2 up and down. As expected, the (p,n) $S$ factor is only sensitive to variations in
the proton width $<\Gamma>_\mathrm{p}$ in the entrance channel, except for an energy
range of up to about 100 keV above the threshold where neutron widths are still small.
This shows that the (p,n) reaction is well suited to study the impact of the optical
potential used to compute the proton widths.
\begin{figure}
\resizebox{\columnwidth}{!}{\rotatebox{270}{\includegraphics[clip=]{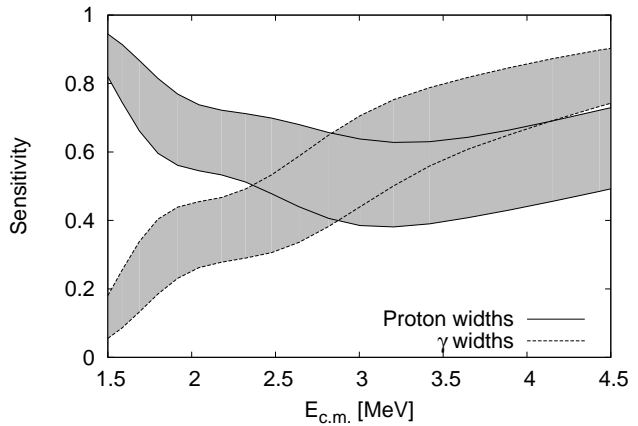}}}
\caption{\label{fig:sensi}Sensitivity of the theoretical $^{70}$Ge(p,$\gamma$)$^{71}$As
results to variations of the widths as function of proton energy. The area between the
solid lines shows the sensitivity to variations of the proton widths, the one between
the dashed lines is the sensitivity to varied $\gamma$ widths.}
\end{figure}

The situation is more complicated in the (p,$\gamma$) case. All widths were varied but
only the proton and $\gamma$ widths were found to be important. The sensitivity to
a variation of the proton and $\gamma$ widths is shown in Fig.\ \ref{fig:sensi},
with the sensitivity $s$ assuming values between Zero and One.  It is defined via
the variation of the cross section ($S$ factor) $\Delta \sigma = s \delta$, with
the width variation factor $\delta$ chosen to be 2 and 0.5, respectively. As can be
seen in Fig.\ \ref{fig:sensi} the $S$ factor is mostly sensitive to the
proton widths at energies below about 2.5 MeV. Above that energy the $\gamma$ width
sensitivity becomes comparable or larger. This implies that -- contrary to the
(p,n) reaction which is sensitive to the proton width at all measured energies --
the $S$ factor is mostly sensitive to the proton width at the lowest energy. The
excellent agreement of theory and measurement shows that the proton width is
described well at low energies and that $\gamma$ widths are described well at the
upper end of the measured energy range.

For completeness, also the impact of a variation of the nuclear level density was
studied. The nuclear level densities in the target and final nucleus determine the
number of possible transitions in the entrance and exit channel, respectively, and thus
impact the size of averaged widths. In the cases studied here, however, a variation
of the nuclear level density does not have any impact. This is because the calculation
uses up to 20 discrete experimental levels in each nucleus at low excitation
energies (see Ref.\ \cite{adndtcs} for a list of the levels). In all cases studied here,
transitions to those lowest levels dominate the widths and therefore varying the
nuclear level density employed above these levels does not change the result.

\subsection{Modification of the optical potential}
\begin{figure}
\resizebox{\columnwidth}{!}{\rotatebox{270}{\includegraphics[clip=]{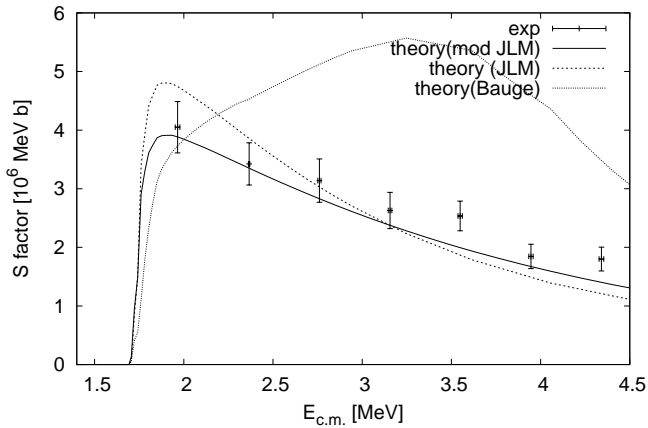}}}
\caption{\label{fig:ge76pn_theory}Astrophysical $S$ factor as function of proton energy
for the reaction $^{76}$Ge(p,n)$^{76}$As. Experimental $S$ factors measured in
this work are compared to NON-SMOKER calculations using the optical potentials of
\cite{jlm,jeuk} (JLM), of \cite{bauge} (Bauge), and a modified JLM potential with
an increased imaginary strength (see text).}
\end{figure}

\begin{figure}
\resizebox{\columnwidth}{!}{\rotatebox{270}{\includegraphics[clip=]{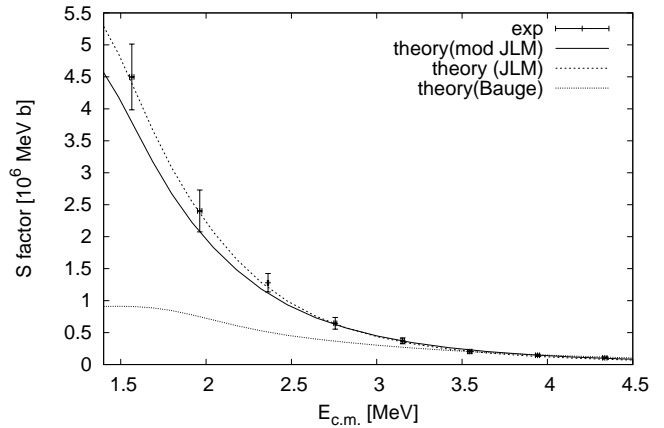}}}
\caption{\label{fig:ge70pg_theory}Astrophysical $S$ factor as function of proton energy
for the reaction $^{70}$Ge(p,$\gamma$)$^{71}$As. Experimental $S$ factors measured in
this work are compared to NON-SMOKER calculations using the optical potentials of
\cite{jlm,jeuk} (JLM), of \cite{bauge} (Bauge), and a modified JLM potential with
an increased imaginary strength (see text).}
\end{figure}

Apparently, the optical potential used for calculation of the proton widths gives
rise to the different energy dependence of the theoretical (p,n) $S$ factor as compared
to experiment. The optical potential used here is the widely-used semi-microscopic
potential of \cite{jlm} with low-energy modifications by \cite{jeuk} (this will be
addressed as JLM potential in the following), based on an infinite nuclear matter
calculation employing the Reid hard core interaction, the Br\"uckner-Hartree-Fock
approximation, and a local density approximation. It has been argued \cite{baugefirst}
that the JLM potential may have to be improved above 160 MeV projectile energy and
that its isovector components may be too weak. The former is not relevant here but
the latter can also have an impact at the low energies of astrophysical interest.
A new parameterization of the JLM model has been derived in \cite{baugefirst} and
then improved to be Lane-consistent, i.e.\ equally applicable to neutrons and protons,
in \cite{bauge}. A comparison of our (p,n) data with the results obtained when using this
latter potential is shown in Fig.\ \ref{fig:ge76pn_theory}. Obviously, the original
JLM potential still fares much better in reproducing both the energy dependence
and the absolute magnitude of the $S$ factor. Similar problems were already found
for Se isotopes in \cite{gyuse}. Fig.\ \ref{fig:ge70pg_theory} shows that the low-energy
$S$ factor of the (p,$\gamma$) reaction is severely underpredicted. The agreement at
the upper end of the measured energy range is not surprising because the sensitivity
to the proton potential is smaller there, as discussed above.

We have attempted to improve the JLM potential by modifying the strengths of its
real and imaginary parts. We find that an increase of the imaginary strength by 70\%
yields the best compromise in reproducing both the (p,n) and (p,$\gamma$) data, as
shown in Figs.\ \ref{fig:ge76pn_theory} and \ref{fig:ge70pg_theory}. The results
are extremely sensitive to a variation of the real part and therefore best agreement
is found when adopting the original real strength. A behavior similar to the
results when using the potential of \cite{bauge} is found for the (p,$\gamma$) reaction when increasing the
real part only slightly.

Since both the isoscalar and isovector parts of the JLM model contribute to the imaginary
part, we cannot disentangle their individual importance. Following the argumentation
of \cite{baugefirst}, we attribute the increase in the imaginary part to a stronger
isovector component. However, the isovector component of the real part has to remain
unchanged. It is conceivable that a different low-energy parameterization of the
JLM model may yield similar results.

\begin{figure}
\resizebox{\columnwidth}{!}{\rotatebox{270}{\includegraphics[clip=]{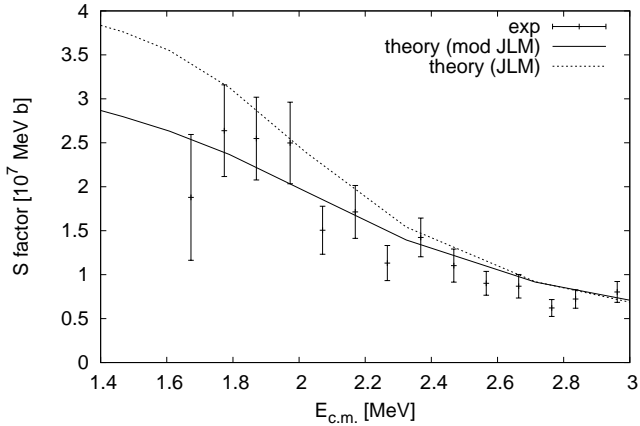}}}
\caption{\label{fig:sr84}Astrophysical $S$ factor as function of proton energy
for the reaction $^{84}$Sr(p,$\gamma$)$^{85}$Y. Experimental $S$ factors from \cite{gyusr}
are compared to NON-SMOKER calculations using
the JLM potential (dashed line) and a modified JLM proton potential (solid line) with
an increased imaginary strength (see text).}
\end{figure}

\begin{figure}
\resizebox{\columnwidth}{!}{\rotatebox{270}{\includegraphics[clip=]{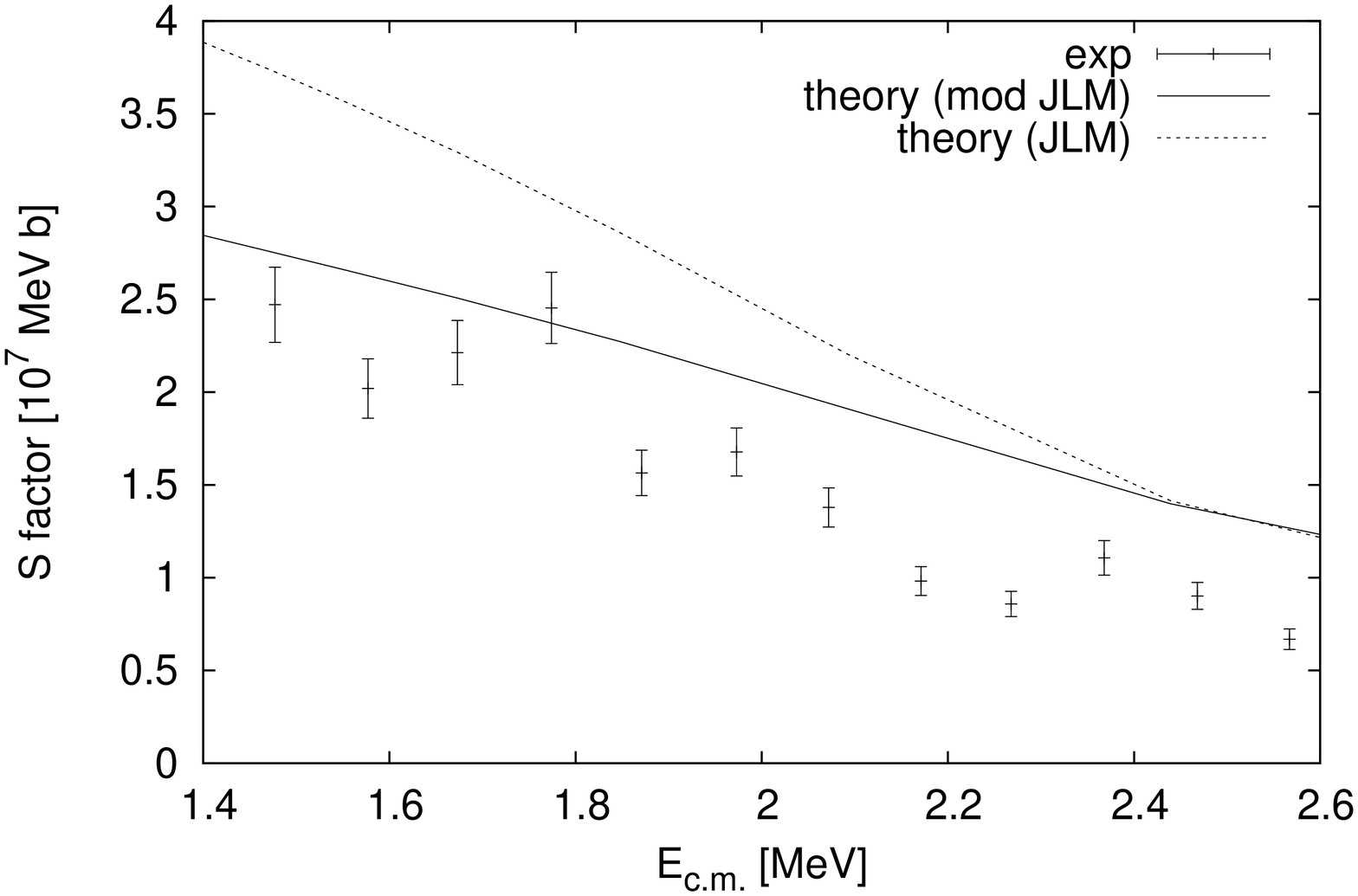}}}
\caption{\label{fig:sr86}Astrophysical $S$ factor as function of proton energy
for the reaction $^{86}$Sr(p,$\gamma$)$^{87}$Y. Experimental $S$ factors from \cite{gyusr}
are compared to NON-SMOKER calculations using the JLM potential (dashed line)
a modified JLM proton potential (solid line) with
an increased imaginary strength (see text).}
\end{figure}

\begin{figure}
\resizebox{\columnwidth}{!}{\rotatebox{270}{\includegraphics[clip=]{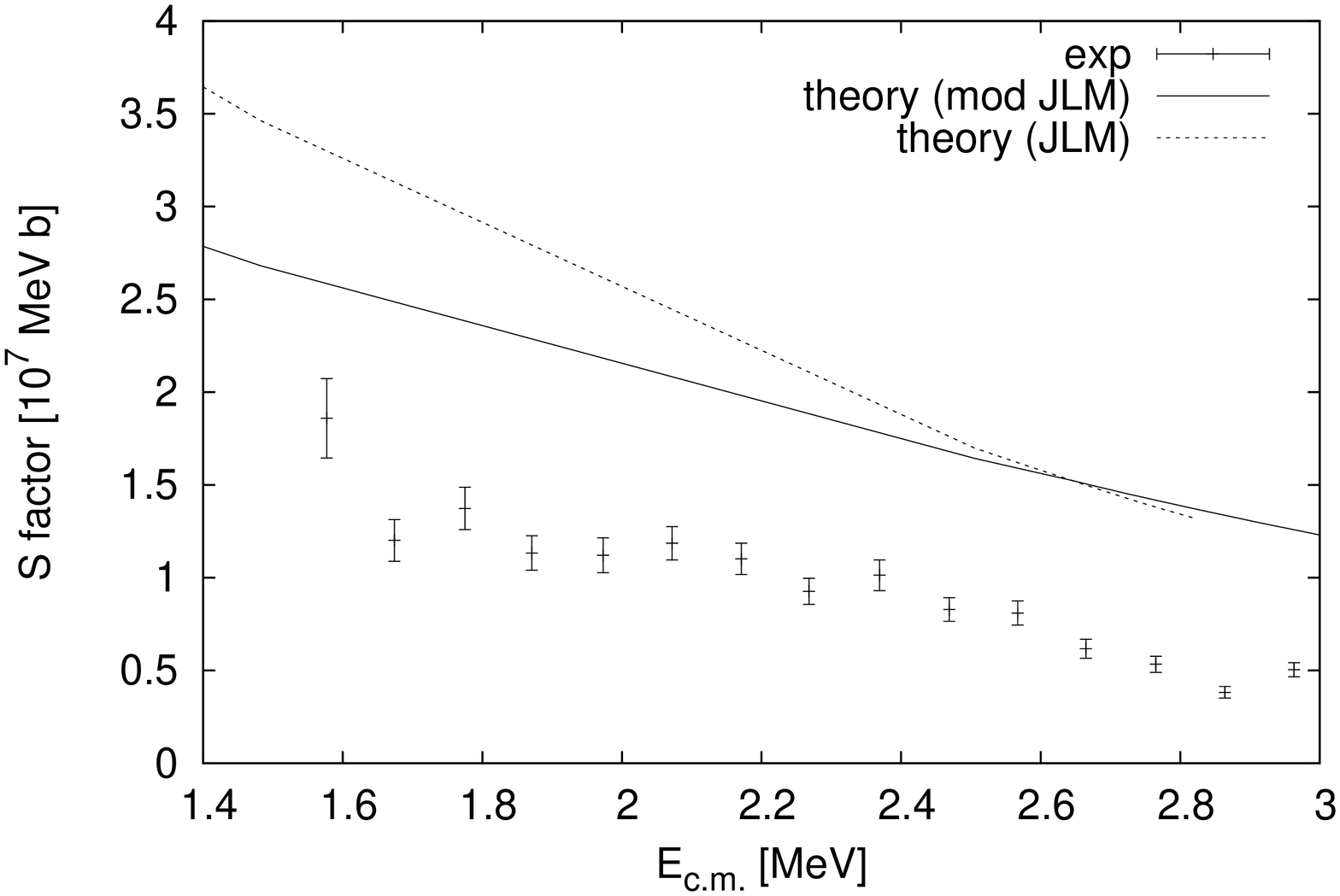}}}
\caption{\label{fig:sr87}Astrophysical $S$ factor as function of proton energy
for the reaction $^{87}$Sr(p,$\gamma$)$^{88}$Y. Experimental $S$ factors from \cite{gyusr}
are compared to NON-SMOKER calculations using the JLM potential (dashed line)
a modified JLM proton potential (solid line) with
an increased imaginary strength (see text).}
\end{figure}

\begin{figure}
\resizebox{\columnwidth}{!}{\rotatebox{270}{\includegraphics[clip=]{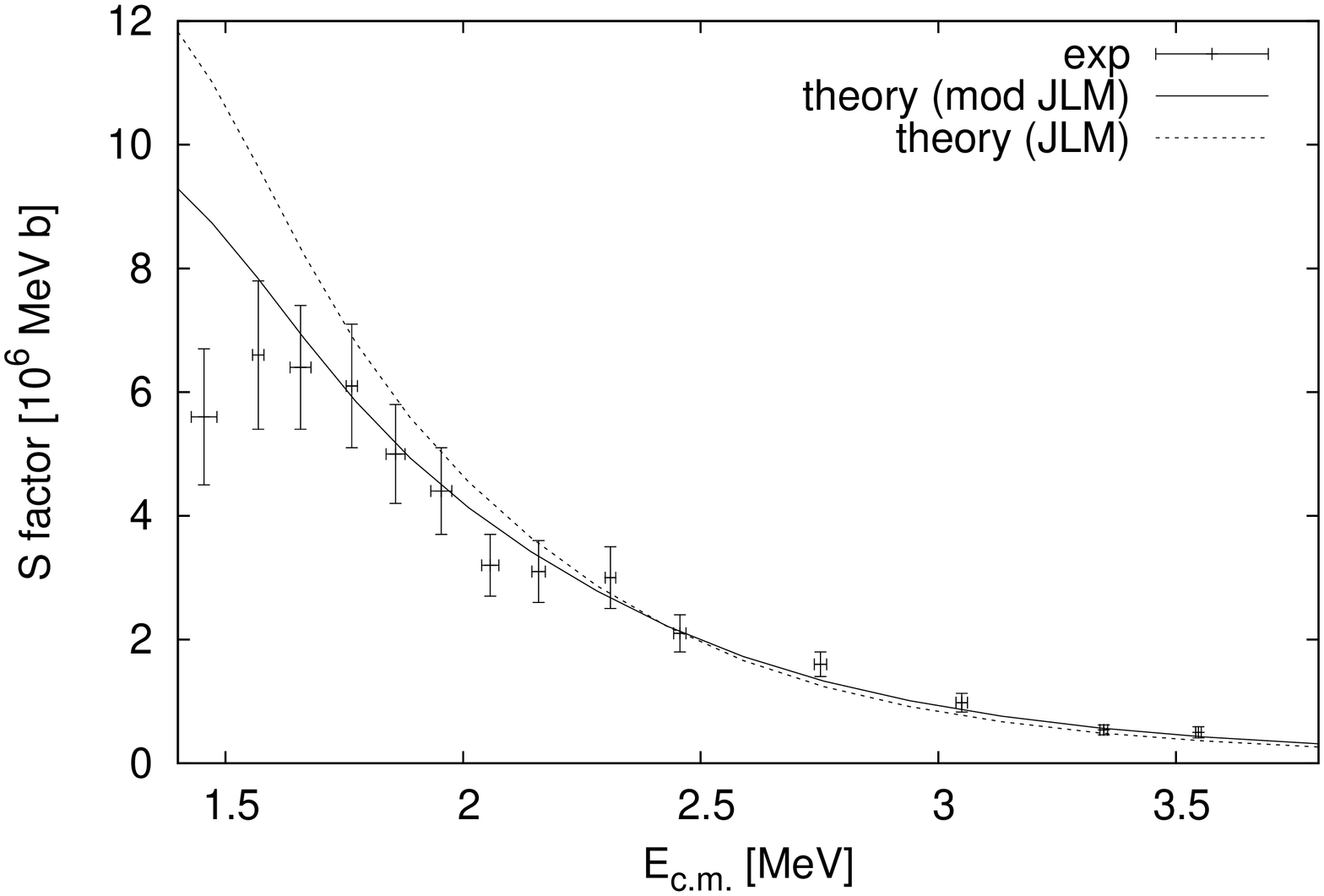}}}
\caption{\label{fig:se74}Astrophysical $S$ factor as function of proton energy
for the reaction $^{74}$Se(p,$\gamma$)$^{75}$Br. Experimental $S$ factors from \cite{gyuse}
are compared to NON-SMOKER calculations using the JLM potential (dashed line)
a modified JLM proton potential (solid line) with
an increased imaginary strength (see text).}
\end{figure}

\begin{figure}
\resizebox{\columnwidth}{!}{\rotatebox{270}{\includegraphics[clip=]{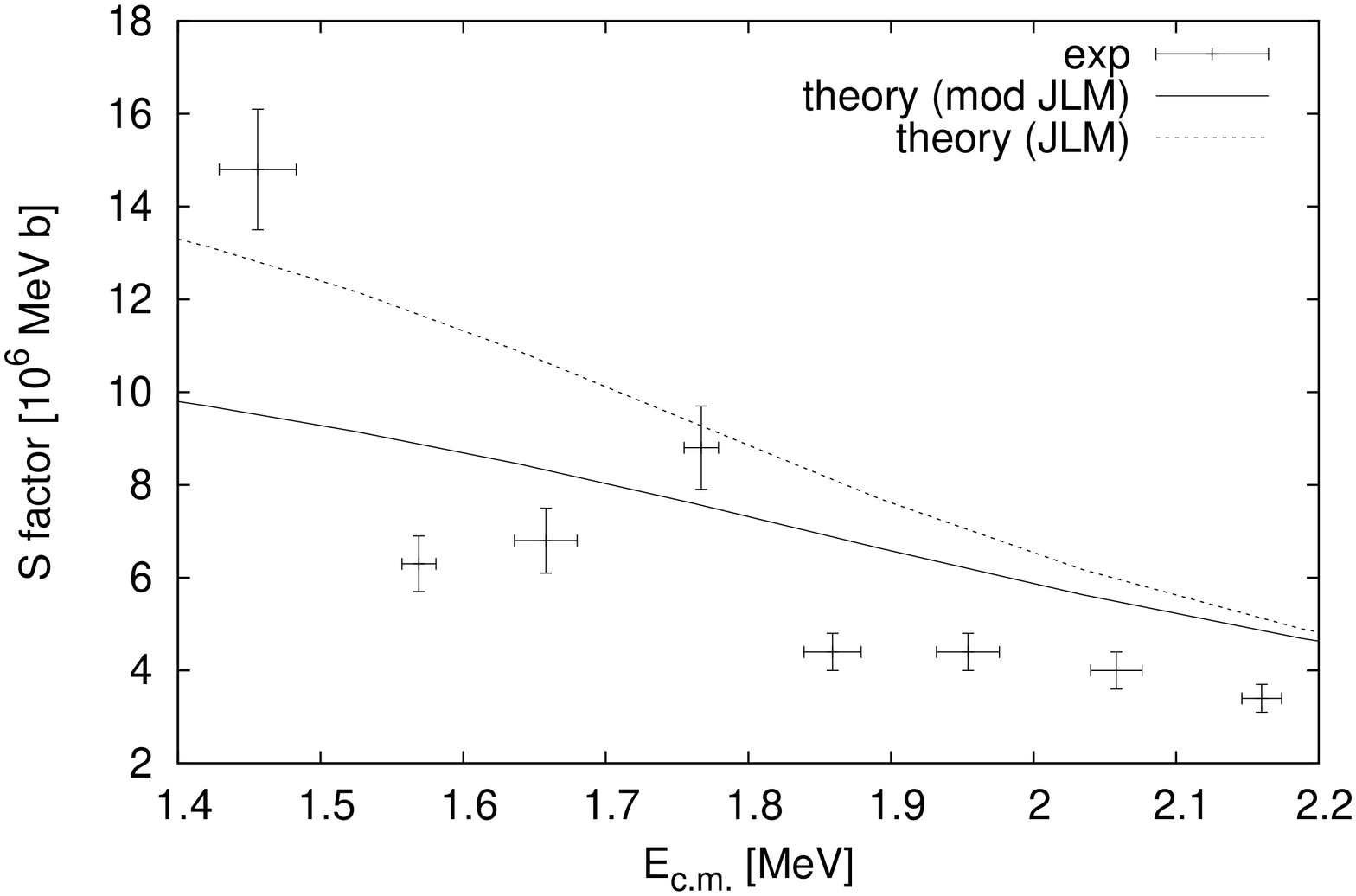}}}
\caption{\label{fig:se76}Astrophysical $S$ factor as function of proton energy
for the reaction $^{76}$Se(p,$\gamma$)$^{77}$Br. Experimental $S$ factors from \cite{gyuse}
are compared to NON-SMOKER calculations using the JLM potential (dashed line)
a modified JLM proton potential (solid line) with
an increased imaginary strength (see text).}
\end{figure}

\begin{figure}
\resizebox{\columnwidth}{!}{\rotatebox{270}{\includegraphics[clip=]{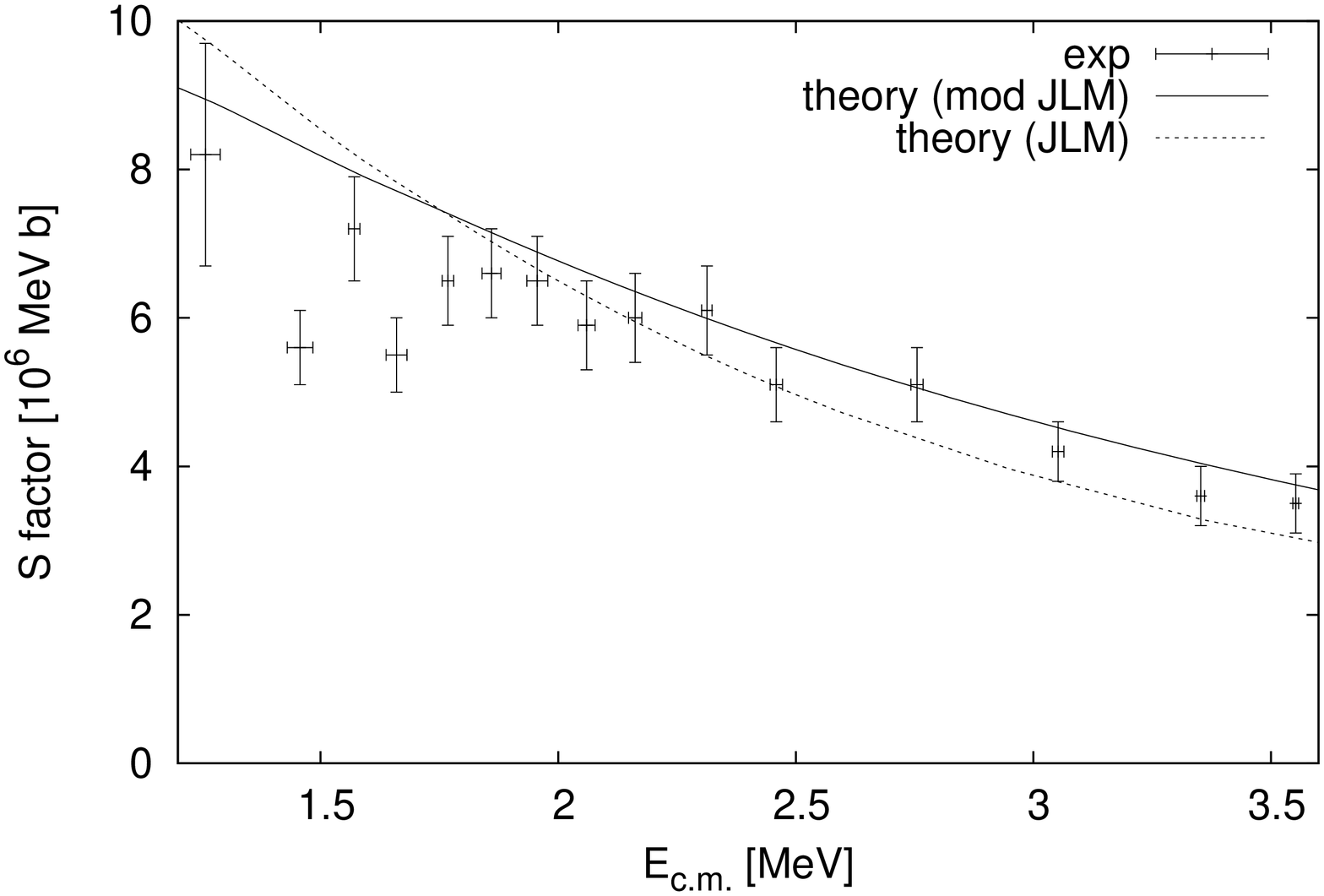}}}
\caption{\label{fig:se82}Astrophysical $S$ factor as function of proton energy
for the reaction $^{82}$Se(p,n)$^{82}$Br. Experimental $S$ factors from \cite{gyuse}
are compared to NON-SMOKER calculations using the JLM potential (dashed line)
a modified JLM proton potential (solid line) with
an increased imaginary strength (see text).}
\end{figure}

Radiative proton capture and (p,n) reactions at astrophysically relevant energies in this
mass region were previously studied involving Sr \cite{gyusr} and Se isotopes \cite{gyuse}.
It is interesting to note that our modified JLM potential also leads to an improved
prediction of the reactions $^{84,86,87}$Sr(p,$\gamma$)$^{85,87,88}$Y,
$^{74,76}$Se(p,$\gamma$)$^{75,77}$Br, and $^{82}$Se(p,n)$^{82}$Br, as shown in Figs.\ \ref{fig:sr84}--\ref{fig:se82} (compare
with NON-SMOKER standard results given in
the original plots of \cite{gyusr,gyuse}). Now all $S$ factors are well
reproduced, except for the one of $^{87}$Sr(p,$\gamma$)$^{88}$Y. Although the theoretical
$S$ factors have come closer to the experimental ones, there still remains a
discrepancy of about a factor of Two. We were not able to establish an unambiguous
explanation for this discrepancy. The predicted
cross sections are insensitive to the theoretical
nuclear level density employed, the level schemes of $^{87}$Sr and $^{88}$Y are
sufficiently well established at low excitation energies, and the deformations of the
nuclei involved are comparable to those of the other cases studied. The only
difference from the experimental point of view is the much longer half-life of
$^{88}$Y (106.7 d). This is longer by a factor of at least 200 than the half-lives
of the other final nuclei appearing in the activation experiments.

It should be mentioned that the JLM approach is also used for neutron potentials.
However, whether the modified imaginary strength also applies to those cannot be
determined within the framework of this paper.

\subsection{Astrophysical reaction rates}

The quantity of primary interest for astrophysical calculations is the
astrophysical reaction rate. The relevant plasma temperature range for classical
$p$ process studies is $2-3$ GK \cite{arnould,rauscher06}. We derived
the reaction rates for the reactions measured in this work directly from the
experimental $S$ factors by linear interpolation and accounting for the
error bars using the code EXP2RATE \cite{exp2rate}.
The numerical integration error is smaller than the experimental uncertainties.
\begin{table}
\caption{\label{tab:ge70rate}Astrophysical reaction rates of the reaction
$^{70}$Ge(p,$\gamma$)$^{71}$As computed from the
experimental data. These are also stellar rates because the stellar enhancement is negligible.}

\begin{tabular}{lr@{$\pm$}l}
\hline \hline
\multicolumn{1}{c}{Temperature}&\multicolumn{2}{c}{Reaction Rate} \\
\multicolumn{1}{c}{[$10^9$ K]}&\multicolumn{2}{c}{[cm$^3$s$^{-1}$mole$^{-1}$]}\\
\hline
2.50&(4.474&0.584)$\times 10^2$\\
2.75&(9.660&1.282)$\times 10^2$\\
3.00&(1.862&0.250)$\times 10^3$\\
3.25&(3.284&0.445)$\times 10^3$\\
3.50&(5.393&0.737)$\times 10^3$\\
3.75&(8.357&1.148)$\times 10^3$\\
4.00&(1.234&0.170)$\times 10^4$\\
4.25&(1.749&0.242)$\times 10^4$\\
4.50&(2.394&0.333)$\times 10^4$\\
4.75&(3.180&0.444)$\times 10^4$\\
5.00&(4.116&0.576)$\times 10^4$\\
\hline \hline
\end{tabular}

\end{table}

Our data allow a computation of the rates for the temperature
interval $(2.5-5.0)\times 10^9$ K in the case of $^{70}$Ge(p,$\gamma$)$^{71}$As.
These rates are given in Table \ref{tab:ge70rate}. Obviously, these are ground state rates because they are derived from experimental data. However, there is no stellar enhancement due to thermal excitation of target states for this reaction in a stellar plasma (see, e.g., \cite{adndt}) and therefore stellar and ground state rates are the same. Because the data excellently confirm the prediction, the use of the tabulated rates in \cite{adndt,adndtcs} is recommended.

Due to the (p,n) threshold at 1.7 MeV, the reaction rate of
$^{76}$Ge(p,n)$^{76}$As can be calculated down to Zero temperature. To get a
better description of the threshold behavior of the $S$ factor, the theoretical
values obtained with the modified JLM potential were used in the rate
calculation below 1.565 MeV proton energy. The resulting rates are listed in
Table \ref{tab:ge76rate}. As in the previous reaction, these are ground state rates. The stellar enhancement predicted \cite{adndt} for this reaction varies smoothly between 4\% ($T_9\leq 0.5$ and $3.0\leq T_9 \leq 4.0$) and 10\% (at $T_9=1.5$). Therefore it is negligible compared to the uncertainties in the rates stemming from the experimental uncertainties. In consequence, the rates quoted in Table \ref{tab:ge76rate} can be taken as both ground state and stellar rates.
Although the rates given in \cite{adndt,adndtcs} were
obtained with the standard JLM potential and the $S$ factors exhibit a slightly
different slope, the new rates still agree reasonably well within errors
with the predicted
ones in the temperature range up to 4 GK.

It is worth to note that the reaction rate of the presently investigated two reactions has been measured by Roughton \textit{et al.} \cite{roughton}. On average the reaction rates obtained for the $^{70}$Ge(p,$\gamma$)$^{71}$As reaction are 35\,\% lower than the presently measured values and 20\,\% higher for $^{76}$Ge(p,n)$^{76}$As (the latter difference being within the error). The reaction rates in \cite{roughton}, however, have been determined from thick target yields and no cross sections have been measured. 

\section{Summary}
We have measured the cross sections of the reactions
$^{70}$Ge(p,$\gamma$)$^{71}$As and $^{76}$Ge(p,n)$^{76}$As by the activation method
and derived the astrophysical $S$ factors in the c.m. energy range of $1.565-4.341$ MeV.
A comparison to predictions from the statistical model showed excellent agreement
with the radiative proton capture data and good agreement with the (p,n) data although
the energy dependence was not accurately reproduced for the latter. A detailed
theoretical study led to the conclusion that the optical potential for protons can
be modified in such a way as to reproduce both the (p,$\gamma$) and (p,n) data. It
proved necessary to increase the depth of the imaginary part of the potential by
a factor of 1.7. Also previously published data for low-energy
(p,$\gamma$) and (p,n) reactions on Se and Sr isotopes
can be reproduced well using this modified
optical potential. It remains to be studied whether the modification is also
applicable to other mass ranges, whether it should be energy dependent, and whether
it should also be applied to neutron potentials. More data
across a wider range of energies and masses would be needed for this.
\begin{table}
\caption{\label{tab:ge76rate}Astrophysical reaction rates of the reaction
$^{76}$Ge(p,n)$^{76}$As computed from the
experimental data. These are also stellar rates because the stellar enhancement is negligible within the given uncertainty.}

\begin{tabular}{lr@{$\pm$}l}
\hline \hline
\multicolumn{1}{c}{Temperature}&\multicolumn{2}{c}{Reaction Rate} \\
\multicolumn{1}{c}{[$10^9$ K]}&\multicolumn{2}{c}{[cm$^3$s$^{-1}$mole$^{-1}$]}\\
\hline
0.25&\multicolumn{2}{c}{0.000}\\
0.50&(3.086&0.012)$\times 10^{-12}$\\
0.75&(2.118&0.040)$\times 10^{-6}$\\
1.00&(1.912&0.071)$\times 10^{-3}$\\
1.25&(1.222&0.066)$\times 10^{-1}$\\
1.50&(2.088&0.142)$\times 10^0$\\
1.75&(1.683&0.133)$\times 10^1$\\
2.00&(8.487&0.746)$\times 10^1$\\
2.25&(3.128&0.296)$\times 10^2$\\
2.50&(9.231&0.919)$\times 10^2$\\
2.75&(2.310&0.239)$\times 10^3$\\
3.00&(5.091&0.542)$\times 10^3$\\
3.25&(1.013&0.110)$\times 10^4$\\
3.50&(1.856&0.205)$\times 10^4$\\
3.75&(3.173&0.355)$\times 10^4$\\
4.00&(5.114&0.580)$\times 10^4$\\
\hline \hline
\end{tabular}

\end{table}

The astrophysical reaction rates derived from the new data for $^{70}$Ge(p,$\gamma$)
and $^{76}$Ge(p,n) are in agreement within the measured temperature range with the
previously published ones \cite{adndt,adndtcs} which were based on Hauser-Feshbach
calculations with the unmodified JLM potential.

\begin{acknowledgments}
This work was supported by OTKA (K68801, T49245, D48283), MTA-OTKA-NSF grant 93/049901,
the Swiss National Science Foundation (grants 200020-061031, 200020-105328) and the
Joint Institute of Nuclear Astrophysics (www.JINAweb.org) NSF-PFC
grant PHY02-16783.
Gy.\ Gy.\ acknowledges financial support from the Bolyai grant.
\end{acknowledgments}

\end{document}